\documentclass[10pt,draft]{dis03}
\usepackage{epsf,amsmath}

\newcommand{\xgam}{%
\ensuremath{%
{x_\gamma}}}
\newcommand{\figref}[1]{Fig.~\ref{#1}}
\newcommand{\pom}{{I\hspace{-0.5ex} P}}
\newcommand{\zpom}{%
\ensuremath{%
{z_\pom}}}
\newcommand{\mjj}{%
\ensuremath{%
{M_{12}}}}
\newcommand{\xpom}{%
\ensuremath{%
{x_\pom}}}

\begin{document}

\title{Diffractive Final States and Tests of QCD
  Factorisation\footnote{Talk presented at XI~Intl. Workshop on Deep-Inelastic Scattering,
DIS 2003, St.~Petersburg.}\ \thanks{This work is supported by the German National
  Merit Foundation.}}
\author{Sebastian Sch\"atzel\thanks{On behalf of the H1 Collaboration.} \\
Physikalisches Institut der Universit\"at Heidelberg \\
69120 Heidelberg, Germany \\
E-mail: schaetzel@physi.uni-heidelberg.de}

\maketitle

\begin{abstract}
\noindent Measurements of hard diffractive final states performed with the H1
experiment at HERA are presented
and confronted with predictions based on diffractive parton densities.
\end{abstract}

\section{Introduction} 
The diffractive structure function of the proton has been measured in
inclusive diffractive deep-inelastic $ep$ scattering (DDIS) by the H1
collaboration~\cite{h1f2d94, h1f2d97, yves}. 
The data are compatible with the decomposition of the structure
function into a flux factor and the structure function of a colourless
exchange (pomeron).
In DGLAP fits, leading order (LO) and next-to-leading order pomeron parton
densities (PPDFs) have 
been extracted and found to be dominated by the gluon distribution~\cite{h1f2d94, h1f2d97, yves}.

\begin{figure}[!thb]
\vspace*{6.0cm} 
\begin{center}
\includegraphics{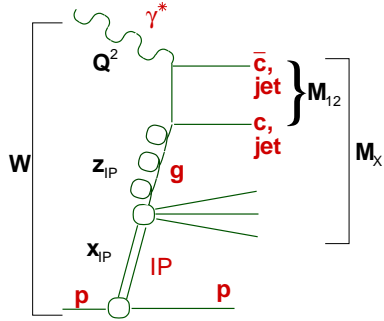}
\vspace*{-2.5cm}
\caption[]{Boson gluon fusion process in DDIS.}
\label{fig:bgf}
\end{center}
\vspace*{-1cm}
\end{figure}

It has been proven within QCD~\cite{collins} that in DDIS the
cross section can be written as a convolution of the diffractive parton
densities of the proton
with the hard parton-photon cross section.
Diffractive dijet and D$^*$ meson (heavy quark) production are
directly sensitive to 
the dominant diffractive gluon through the boson gluon fusion production mechanism
(\figref{fig:bgf}) and are used to test
QCD hard scattering factorisation in diffraction.
A photon of 
  virtuality $Q^2$ undergoes a hard scatter with
a diffractive gluon forming a $q\bar{q}$ pair. The centre-of-mass
energy of the hard scattering process is labelled \mjj.
The gluon carries a fraction \zpom{} of the pomeron momentum. The
pomeron carries a fraction \xpom{} of the proton momentum.
The $\gamma p$ and pomeron-proton centre-of-mass energies are denoted
by $W$ and $M_X$, respectively.
The data are compared to the LO predictions of the Monte Carlo
program RAPGAP~\cite{rapgap}. Higher-order effects are modelled by using parton showers.
The predictions are based on the pomeron model~\cite{rpm} with flux factor and LO PPDFs
as extracted in inclusive DDIS. The PPDFs from the `H1 fit
2'~\cite{h1f2d94} and the new 
`H1 2002 fit'~\cite{h1f2d97} to more precise data are used.

\section{Hard Final States in Diffractive Deep-Inelastic Scattering}
\begin{figure}[!thb]
\vspace*{4.5cm} 
\begin{center}
\includegraphics{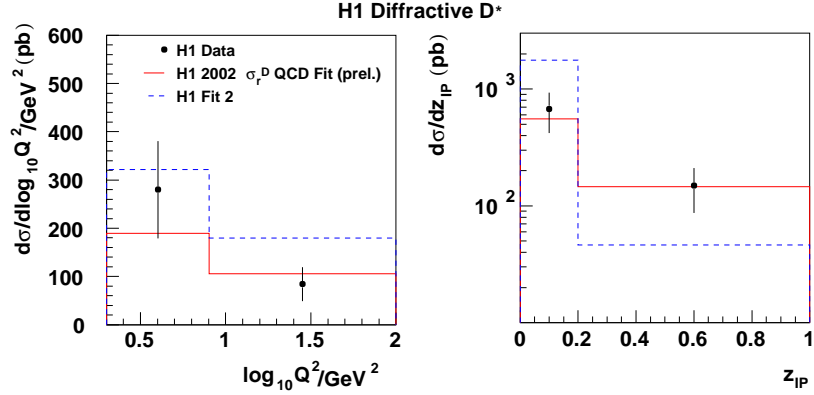}
\caption[]{Differential cross sections for D$^*$ production in DDIS.}
\label{fig:dstar}
\end{center}
\vspace*{0cm}
\end{figure}

\begin{figure}[!thb]
\vspace*{4.5cm} 
\begin{center}
\includegraphics{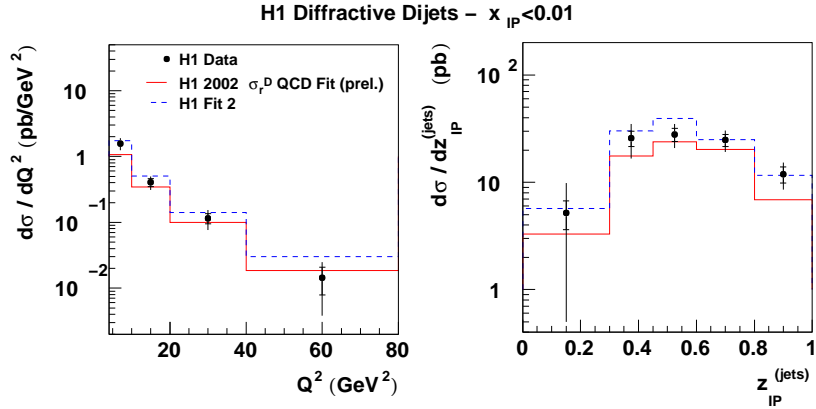}
\caption[]{Differential cross sections for dijet production in
  DDIS.}
\label{fig:disjets}
\end{center}
\vspace*{-1cm}
\end{figure}

In \figref{fig:dstar}, differential cross sections for D$^*$
production in DDIS ($2<Q^2<100$~GeV$^2$,
$\xpom<0.04$, $E_T^{\rm D^*}>2$~GeV)
are shown~\cite{h1f2d97,disdstar}.
The new `H1 2002 fit' based prediction describes the data well.
The `H1 fit 2' predicts a $\approx$25\% larger rate. This difference
is of the order of the uncertainty arising through the precision with which the gluon density is known.
Differential cross sections for dijet production in DDIS ($4<Q^2<80$ GeV$^2$,
$\xpom<0.01$, $E_T^{\rm jet1,2}>4$~GeV)~\cite{h1f2d97,disjets} are
shown in \figref{fig:disjets}. Both predictions give a good
description of the shapes and normalisation of the data.
The predictions based on PPDFs extracted in inclusive DDIS describe 
DDIS dijet and D* 
production within the uncertainties of the PPDFs.
At the present level of accuracy, QCD hard scattering
factorisation holds in DDIS.

\section{Dijets in Diffractive Photoproduction}
\begin{figure}[!thb]
\vspace*{4.5cm} 
\begin{center}
\includegraphics{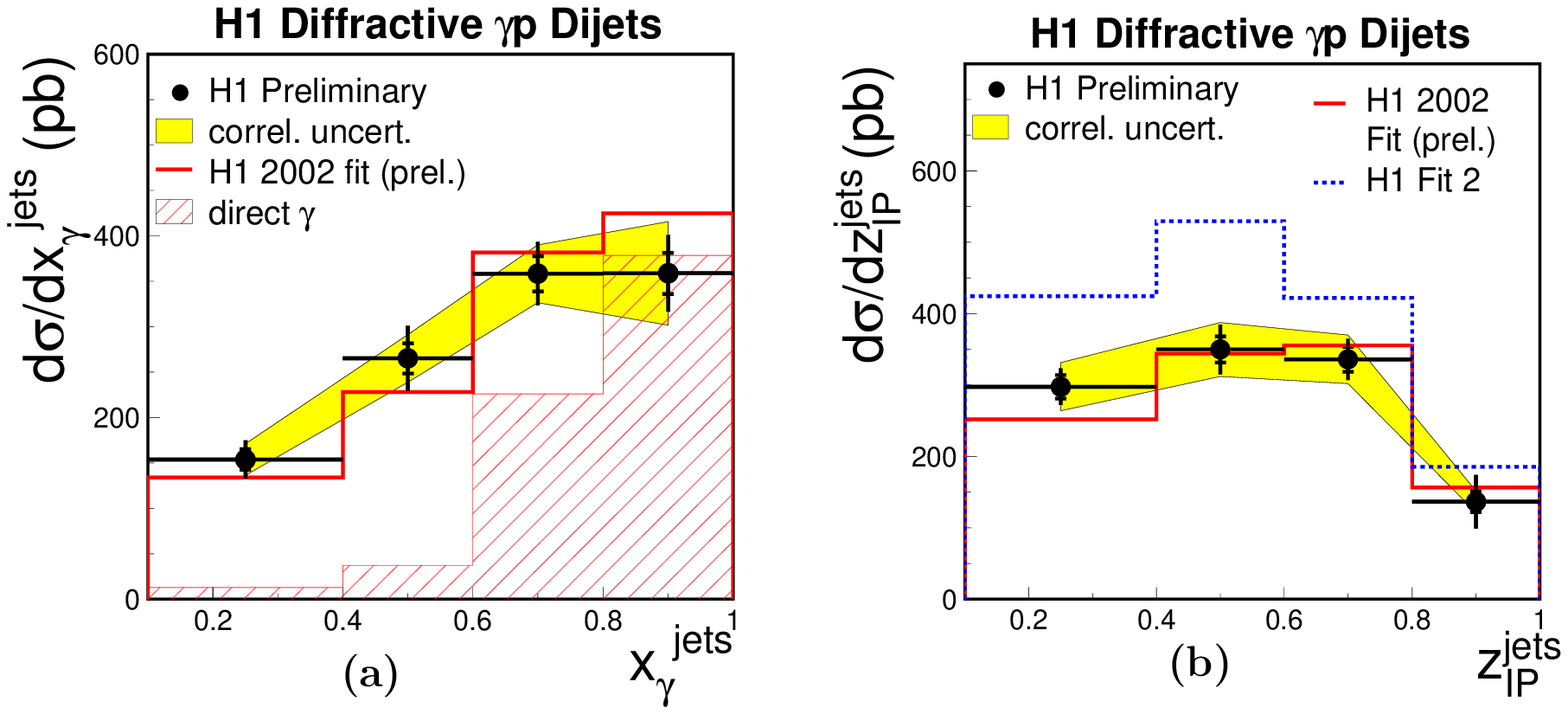}
\caption[]{Differential cross sections for diffractive dijet photoproduction.}
\label{fig:xgamzpom}
\end{center}
\vspace*{-0.5cm}
\end{figure}

\begin{figure}[!thb]
\vspace*{7cm} 
\vspace*{-0.5cm} 
\begin{center}
\includegraphics{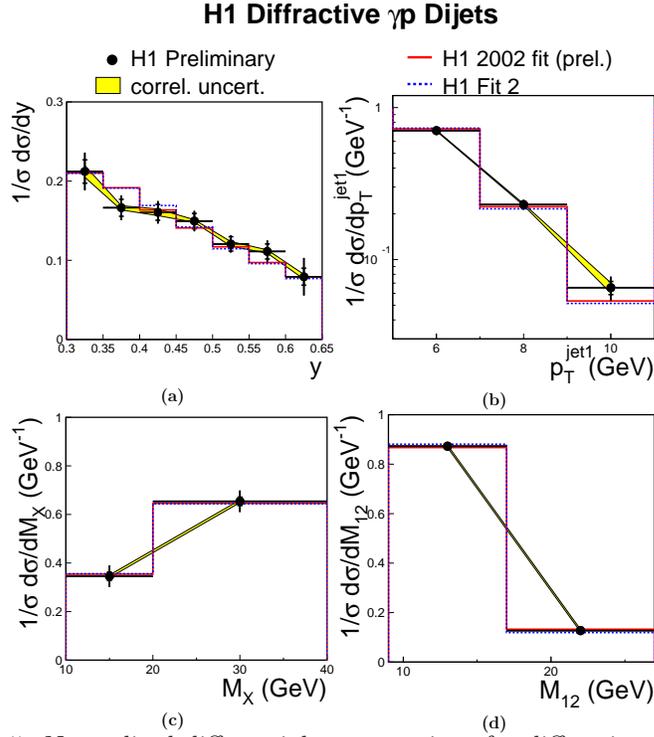}
\vspace*{2cm}
\caption[]{Normalised differential cross sections for diffractive dijet
  photoproduction.}
\label{fig:normxs}
\end{center}
\vspace*{-1cm}
\end{figure}

In diffractive dijet photoproduction ($Q^2<0.01$ GeV$^2$,
$\xpom<0.03$, $E_T^{\rm jet1}>5$~GeV, $E_T^{\rm jet2}>4$~GeV) at
HERA~\cite{eps03}, the quasi-real photon can fluctuate
into a hadronic system of which a parton with momentum
fraction \xgam{} undergoes the hard scatter (`resolved photon'
process). In `direct photon' processes  (\xgam=1), the photon itself
enters the hard scatter.
The cross section for dijet production is shown in
\figref{fig:xgamzpom}a\footnote{In the comparisons shown at the DIS'03 conference, incorrect
  values for $\alpha_s$ were used for the predictions, resulting
  in dijet cross sections which were too large by a factor $\approx$1.4.} as a function of \xgam. 
The prediction based on the `H1 2002 fit' gives a good description
both in shape and normalisation throughout the \xgam{} range.
The cross section is shown in \figref{fig:xgamzpom}b as a function of \zpom.
The `H1 fit 2' prediction overestimates the normalisation of the data
by a factor $\approx$1.4. The difference of the predictions is of the
order of the uncertainty arising from that of the gluon distribution. Both predictions give good descriptions of
the measured shape. Normalised differential cross sections in other
characteristic variables are shown in \figref{fig:normxs}. The variable $y$ relates the $ep$ and $\gamma p$  centre-of-mass energies $\sqrt{s}$ and $W$ via $W=\sqrt{y\,s}$.
All shapes
are described by the predictions.

Using the pomeron model to relate the measurements, diffractive dijet
cross sections in DIS and photoproduction can be compared.
In  photoproduction a suppression 
relative to DIS is found of 1.3$\pm$0.3 (exp.), where the uncertainty
is estimated from the total experimental errors of both measurements
only. The factor is independent of the PPDFs used.
The suppression is not significant at the 
present level of precision and there
is no evidence that it differs between direct and
resolved photon processes.

\section{Conclusions}
In diffractive DIS, measured dijet and D$^*$ production cross sections are in agreement with predictions which rely on QCD factorisation.
For diffractive
dijet photoproduction, an overall suppression factor relative to DIS
 of 1.3$\pm$0.3
(exp.) is found which does not differ between direct and resolved
photon processes.
 The diffractive $\gamma p$ dijet cross sections are
compatible with the predictions within the relatively large
uncertainties of the data.
This is in contrast to the situation in hadron-hadron collisions at
Fermilab where a large suppression of the single-diffractive dijet 
cross section relative
to predictions using DIS pomeron parton densities is
observed.

\end{document}